
\magnification=1200
\baselineskip=13pt
\overfullrule=0pt
\tolerance=100000

\rightline{MRI-Phy/20/95}
\rightline{UR-1437\ \ \ \ \ \ \ }
\rightline{hep-th 9510075}

\bigskip

\baselineskip=18pt

\centerline{\bf Gelfand-Dikii Brackets for Nonstandard
               Supersymmetric Systems}

\bigskip
\bigskip

\centerline{Ashok Das{\footnote*{e-mail: das@urhep.pas.rochester.edu}}}
\centerline{Department of Physics and Astronomy}
\centerline{University of Rochester}
\centerline{Rochester, NY 14627}
\centerline{and}
\centerline{Sudhakar Panda{\footnote{**}{e-mail: panda@mri.ernet.in}}}
\centerline{Mehta Research Institute of Mathematics}
\centerline{and Mathematical Physics}
\centerline{10, Kasturba Gandhi Marg}
\centerline{Allahabad 211002, India}

\bigskip
\bigskip
\bigskip
\bigskip

\centerline{\bf Abstract}

We show how a general nonstandard Lax equation (supersymmetric or otherwise)
can be expressed as a
standard Lax equation. This enables us to define the Gelfand-Dikii brackets for
a nonstandard supersymmetric equation. We discuss the Hamiltonian structures
for the nonstandard super KP system  and work out
explicitly the two Hamiltonian structures of the supersymmetric Two Boson
system from this point of view.

\medskip

\vskip 2 truein

\vfill\eject
\noindent{\bf 1. Introduction:}
\medskip
A Lax description of integrable systems is quite useful from many points of
view. Most integrable systems can be described by a standard Lax equation [1-3]
in terms of an appropriate pseudo-differential operator. More recently,
however,
there has been considerable interest in the Two Boson (TB) equation [4-5] and
its supersymmetric version [6-8] which have a nonstandard Lax representation.
Namely, if $L$ denotes the appropriate pseudo-differential operator
(Lax operator), then these equations can be written as

$$  {\partial L\over \partial t} = [ L , (L^{2})_{\geq 1} ]\eqno(1)$$
While the conserved quantities for such systems continue to be defined as ${\rm
Tr} L^{n}$, it is not clear whether the standard Gelfand-Dikii brackets will
continue to define the Hamiltonian structures for such nonstandard systems. In
fact, a modification of the Gelfand-Dikii brackets in the case of the Two
Boson equation was already noted in ref.[9] and it is the purpose of this
letter
to extend such a generalization to superspace to include supersymmetric
nonstandard integrable systems. The Hamiltonian structures for the
supersymmetric Two Boson (sTB) have, of course, already been obtained directly
[6-7].
However, a Gelfand-Dikii description of these structures has proven elusive so
far. In this letter, we define the Gelfand-Dikii brackets for this system which
leads to the correct Hamiltonian structures. Our results are organized as
follows. In section 2, we show how a nonstandard Lax system can be expressed
as a standard system which would allow us to define the Gelfand-Dikii brackets
in the conventional manner. We discuss the Hamiltonian structures of the
nonstandard super KP system [7] from this point of view. In section 3, we
construct
explicitly the two Hamiltonian structures associated with the supersymmetric
Two Boson (sTB) system. We end with a brief conclusion in section 4.
\medskip
\noindent{\bf 2. Transformation to a Standard System:}
\medskip
Let us consider a pseudo-differential operator of the form [3,10]

$$  L = \partial + \sum_{i=0}^{\infty} u_{i-1}\partial^{-i}\eqno(2)$$
where $u_{i}$'s define the dynamical variables with the dynamics given by the
nonstandard Lax equation

$$ {\partial L\over \partial t} = [ L , (L^{2})_{\geq 1} ]\eqno(3)$$
Throughout our discussions, $(A)_{\geq 1}$ would stand for the purely
differential part of a pseudo-differential operator while $(A)_{+}$ would
correspond to the part of the pseudo-differential operator without any negative
power. (In other words, it will include the nonderivative term as well as the
differential part.)

Let us next make the transformation

$$ {\cal D} = e^{-\omega}\partial e^{\omega}\equiv (\partial+(\partial\omega))
 \eqno(4)$$
where we assume that $\omega$ is a bosonic function. Under this transformation,
we can write

$$ {\cal L} = e^{-\omega} L e^{\omega}\equiv {\cal D} + \sum_{i=0}^{\infty}
u_{i-1}{\cal D}^{-i} \eqno(5)$$
In other words, with this transformation, pseudo-differential operators are
expanded in a basis in powers of ${\cal D}$.
It is straightforward to check now that eq. (3) reduces, under the
transformations in eqs. (4-5) to a standard equation of the form

$$ {\partial {\cal L}\over \partial t} = [ {\cal L} , ({\cal L}^{2})_{+}
]\eqno(6)$$
By construction, however, the right hand side of eq. (6) is a
pseudo-differential operator of negative powers whereas the left hand side is
not unless

$$ (\partial\omega) = -u_{-1}\eqno(7)$$
Thus we see that a nonstandard Lax equation can be rewritten as a standard Lax
equation through a transformation of the form in eq. (4) provided the parameter
of the transformation is related to the coefficient of $\partial^{0}$ of the
Lax operator as in eq. (7). Such an observation, in fact, had led to the
definition of
the Gelfand-Dikii brackets in the case of the TB system [9]. However, the
present
derivation  is quite general and holds even in super space since we have not
assumed any specific property of $\partial$. (As is clear from this discussion,
this works only if the coefficient of $\partial^{0}$ is nonzero. It remains an
open question as to whether a nonstandard equation with this term vanishing can
be put into a standard form.)

Let us next discuss the general features of the Hamiltonian structures for the
nonstandard super KP system based on our results. Let us consider a bosonic
pseudo-differential operator in superspace of the form [7]

$$ L = D^{2} + \sum_{i=0}^{\infty} \Psi_{i-1} D^{-i}\eqno(8)$$
where the superspace covariant derivative is defined to be

$$ D\equiv {\partial\over \partial\theta} + \theta {\partial\over\partial
x}\eqno(9)$$
and the coefficient functions, $\Psi_{i}$, are assumed to be superfields with a
Grasmannian grading $|i|$. The nonstandard super KP equations are obtained from
the nonstandard Lax equation

$$ {\partial L\over \partial t_{n}} = [ L , (L^{n})_{\geq 1} ]\eqno(10)$$
where $n=1,2,3,\cdots$.

Following our earlier discussion, we  define

$$ \eqalign{{\cal D} & = e^{-\omega} D e^{\omega} = (D + (D\omega))\cr
{\cal L} & = e^{-\omega} L e^{\omega} = {\cal D}^{2} + \sum_{i=0}^{\infty}
\Psi_{i-1} {\cal D}^{-i}}\eqno(11)$$
where we assume that $\omega$ is a bosonic superfield satisfying (see eq. (7))

$$ (D^{2}\omega) = -\Psi_{-1}\eqno(12)$$
The dynamical equations for the nonstandard sKP equations can now be obtained
from the standard Lax equation

$$ {\partial {\cal L}\over \partial t_{n}} = [ {\cal L} , ({\cal L}^{n})_{+} ]
\eqno(13)$$
We can now define the dual operators as

$$\eqalign{{\cal Q} & = \sum_{i=0}^{\infty}{\cal D}^{i-1} q_{i-1}\cr
  {\cal V} & = \sum_{i=0}^{\infty}{\cal D}^{i-1} v_{i-1}}\eqno(14)$$
which we assume to be bosonic so that the linear functional can be defined to
be

$$ F_{\cal Q}({\cal L}) = {\rm sTr} ( {\cal L}{\cal Q} )\eqno(15)$$
where the supertrace (sTr) is defined as an integral over the superspace of the
super residue which is defined as the coefficient function of ${\cal D}^{-1}$
with the ${\cal D}^{-1}$ operator on the right.

Since the transformed Lax equation is a standard equation, we can define the
Gelfand-Dikii brackets in the conventional manner from [11]

$$\eqalign{\{F_{{\cal Q}}({\cal L}),F_{{\cal V}}({\cal L})\}_{1} & = -{\rm
sTr}({\cal L}[{\cal Q},{\cal V}]) \cr
\{F_{{\cal Q}}({\cal L}),F_{{\cal V}}({\cal L})\}_{2} & = -{\rm sTr}[(({\cal
L}{\cal V})_{+}{\cal L}-{\cal L}({\cal V}{\cal L})_{+}){\cal Q}]}\eqno(16)$$
Using the inverse transformation (see eq. (11)),

$$\eqalign{ L & = e^{\omega}{\cal L}e^{-\omega}\cr
            Q & = e^{\omega}{\cal Q}e^{-\omega}\cr
            V & = e^{\omega}{\cal V}e^{-\omega}}$$
we can then, see that the
Gelfand-Dikii brackets, in terms of the original variables, take the form

$$\eqalign{\{F_{Q}(L),F_{V}(L)\}_{1} & = -{\rm sTr}(L[Q,V]) \cr
 \{F_{Q}(L),F_{V}(L)\}_{2} & = -{\rm sTr}[((LV)_{+}L - L(VL)_{+})Q]}\eqno(17)$$
which would explain the form of the brackets used in ref.[12] to obtain the
Hamiltonian structures. Since these structures have already been worked out, we
donot go any further into details of this calculation.
\medskip
\noindent{\bf 3. Hamiltonian Structures for sTB System:}
\medskip
The Lax operator for the sTB equation is given by

$$ L = D^{2}-(D\Phi_{0})+D^{-1}\Phi_{1}\eqno(18)$$
where $D$ is the covariant derivative in superspace defined in eq. (9) and
$\Phi_{0}$, $\Phi_{1}$ are two fermionic superfields. The dynamical equations
are obtained from the nonstandard Lax equation (One can, of course, write the
equation for the hierarchy as well, but we are concentrating on the sTB
equation for simplicity.)

$$ {\partial L\over \partial t} = [L , (L^{2})_{\geq 1} ]\eqno(19)$$

Following our earlier discussion, we define

$$\eqalign{{\cal D} & = e^{-\omega} D e^{\omega} = (D+(D\omega)) \cr
 {\cal L} & = e^{-\omega} L e^{\omega} = {\cal D}^{2}-(D\Phi_{0})+{\cal
D}^{-1}\Phi_{1} }\eqno(20)$$
where

$$ (D^{2}\omega) = (D\Phi_{0})\eqno(21)$$
With these transformations, the sTB equations can be easily checked to result
from the standard Lax equation

$$ {\partial {\cal L}\over\partial t} = [{\cal L} , ({\cal L}^{2})_{+}
]\eqno(22)$$
In this case, then, we can define the bosonic dual operators as

$$\eqalign{{\cal Q} & = q_{0} + {\cal D}^{-1}q_{1}\cr
 {\cal V} & = v_{0} + {\cal D}^{-1}v_{1}}\eqno(23)$$
and the linear functional

$$F_{{\cal Q}}({\cal L}) = {\rm sTr}({\cal L}{\cal Q}) = \int dx\;d\theta
((D\Phi_{0})q_{1}-\Phi_{1}q_{0})\eqno(24)$$
This allows us to obtain the first Hamiltonian structure from the standard
definition

$$ \{F_{{\cal Q}}({\cal L}), F_{{\cal V}}({\cal L})\}_{1} = -{\rm sTr}({\cal
L}[{\cal Q}, {\cal V}])\eqno(25)$$
which through an inverse transformation also gives (The additional negative
sign in the case of a supersymmetric system, compared to the bosonic one,
results from our definition of the "sRes".)

$$ \{F_{Q}(L), F_{V}(L)\}_{1} = -{\rm sTr}(L[Q,V])\eqno(26)$$
Explicitly, this gives the first Hamiltonian structure (as an operator) for the
sTB system to be

$$ {\cal D}_{1}^{\rm sTB} = \left(\matrix{0 & -D\cr
                                          -D & 0}\right)\eqno(27)$$
which is the correct structure obtained in ref.[6-7].

To obtain the second Hamiltonian structure, we note that the consistency of the
Lax equation in (22) requires that the right hand side should only have terms
with negative powers of ${\cal D}$. This makes it a constrained system and one
has to be careful in defining duals and Hamiltonian structures in such a case.
As is described in ref.[10] (with suitable generalization to superspace), a
consistent definition of the dual in this case would be so as to yield

$$ {\rm sRes}[{\cal L},{\cal V}] = 0\eqno(28)$$
It is easy to check that with the dual defined in eq. (23), this condition is
not satisfied. Consequently, we have to modify the definition of the dual which
we take to be

$${\bar{\cal V}} = v_{-2}{\cal D}^{2} + {\cal V}\eqno(29)$$
and similarly for ${\bar{\cal Q}}$. (We parenthetically remark here that we
could have chosen the modification to be a term linear in ${\cal D}$ as well
which, however, doesnot change the result.) Furthermore, the modification of
the dual does
not change the linear functional defined in eq. (24). Requiring eq. (28)
to hold, namely,

$$ {\rm sRes}[{\cal L},{\bar{\cal V}}] = 0\eqno(30) $$
determines the coefficient $v_{-2}$ as

$$ \Phi_{1}v_{-2} = v_{1}\eqno(31)$$
and similarly for $q_{-2}$.

The second Hamiltonian structure can now be defined in a straightforward
manner [11] as follows.

$$ \eqalign{\{F_{{\bar{\cal Q}}}({\cal L}),F_{{\bar{\cal V}}}({\cal L})\}_{2} &
= \{F_{{\cal Q}}({\cal L}),F_{{\cal V}}({\cal L})\}_{2}\cr
 & = -{\rm sTr}[(({\cal L}{\bar{\cal V}})_{+}{\cal L}-{\cal L}({\bar{\cal
V}}{\cal L})_{+}){\bar{\cal Q}}]}\eqno(32)$$
This can also be expressed more elegantly in terms of the duals ${\cal Q}$ and
${\cal V}$ as

$$ \eqalign{\{F_{{\cal Q}}({\cal L}),F_{{\cal V}}({\cal L})\}_{2} & = -{\rm
sTr}[(({\cal L}{\cal V})_{+}{\cal L}-{\cal L}({\cal V}{\cal L})_{+})
{\cal Q}]\cr
 & + \int dx\;d\theta\;(D^{-1}{\rm sRes}[{\cal Q},{\cal L}])({\rm sRes}[{\cal
V},{\cal L}])\cr
 & - \int dx\;d\theta\;[({\rm sRes}[{\cal Q},{\cal L}])({\rm sRes}{\cal L}{\cal
V}{\cal D}^{-1})-({\rm sRes}[{\cal V},{\cal L}])({\rm sRes}{\cal L}{\cal
Q}{\cal D}^{-1})]}\eqno(33)$$
In terms of the original variables, the second Hamiltonian structure can, then,
be obtained from

$$\eqalign{\{F_{Q}(L),F_{V}(L)\}_{2} & = -{\rm sTr}[((LV)_{+}L-L(VL)_{+})Q]\cr
 & + \int dx\;d\theta\;(D^{-1}{\rm sRes}[Q,L])({\rm sRes}[V,L])\cr
 & - \int dx\;d\theta\;[({\rm sRes}[Q,L])({\rm
sRes}LVD^{-1})-({\rm sRes}[V,L])({\rm sRes}LQD^{-1})]}\eqno(34)$$
We observe here that the generalization of the superspace Gelfand-Dikii
brackets in the present case is quite analogous to what was obtained for the TB
system. (We correct here some misprints in the definition of the Gelfand-Dikii
bracket in ref. [9]. The correct bracket should be

$$\eqalign{\{F_{Q}(L),F_{V}(L)\} & = {\rm Tr} [((LV)_{+}L-L(VL)_{+})Q]\cr
              & + \int dx\;(\partial^{-1}{\rm Res}[Q,L])({\rm Res}[V,L])\cr
& + \int dx\;[({\rm Res}[Q,L])({\rm Res}LV\partial^{-1})-({\rm Res}[V,L])({\rm
Res}LQ\partial^{-1})]}
$$
The structure in eq. (34) can now be compared with this.)
Without going into a lot of tedious algebra, we simply note here that
eqs. (33-34) lead explicitly to the second Hamiltonian structure of the sTB
system (as an operator) of the form

$$ {\cal D}_{2}^{\rm sTB} =
\left(\matrix{-2D-2D^{-1}\Phi_{1}D^{-1}+D^{-1}(D^{2}\Phi_{0})D^{-1} &
D^{3}-D(D\Phi_{0})+D^{-1}\Phi_{1}D\cr
-D^{3}-(D\Phi_{0})D-D\Phi_{1}D^{-1} & -(D^{2}\Phi_{1}+\Phi_{1}D^{2})}\right)
\eqno(35)$$
This is, indeed, the correct Hamiltonian structure that was derived in
ref.[6-7].
In spite of its complicated structure, it is easy to see that it has the right
symmetry properties and it can be shown through the methods of
super-prolongation that this satisfies the super- Jacobi identity.

Finally, we would like to make some comments on the constraint in eq. (30)
since it has not been properly emphasized in the literature. We note from eqs.
(20-21) that the transformed basis of (super) pseudo-differential operators,
${\cal D}$, contain dynamical variables and, therefore, evolve with time. As a
result, we have

$$ {\partial F_{{\bar{\cal Q}}}({\cal L})\over \partial t} = {\rm
sTr}({\partial {\cal L}\over\partial t}{\bar{\cal Q}}+{\cal L}{\partial
{\bar{\cal Q}}\over\partial t})\eqno(36)$$
It is no longer true, therefore, that

$${\partial F_{{\bar{\cal Q}}}({\cal L})\over \partial t} = {\rm sTr}({\partial
{\cal L}\over \partial t}{\bar{\cal Q}}) = {\rm sTr}([{\cal L}, ({\cal
L}^{2})_{+}]{\bar{\cal Q}})\eqno(37)$$
which is crucial for an analysis of the Hamiltonian flow of the dynamical
equations. With the constraint, eq. (30), however, it can be checked that eq.
(37) holds and the Hamiltonian analysis carries through.
\vfill\eject

\noindent{\bf 4. Conclusion:}
\medskip
In this letter, we have tried to define the Gelfand-Dikii brackets for
nonstandard supersymmetric integrable systems. We have shown that a nonstandard
Lax equation (in ordinary space or in a superspace) can be brought under
suitable conditions to a standard Lax equation through an appropriate
transformation. This, then, allows us to define the Gelfand-Dikii brackets in a
conventional manner in the transformed variables. This can be transformed back
to the original variables in a straightforward manner. We have discussed the
structure of the Gelfand-Dikii brackets of the supersymmetric nonstandard KP
system from this point of view. We have also explicitly constructed the
Hamiltonian structures of the sTB system from these Gelfand-Dikii brackets. In
this case, nontrivial constraints arise which must be taken care of and we
discuss the meaning of these constraints.

One of us (A.D.) would like to thank Prof. H. Mani for the hospitality at the
Mehta Research Institute where part of this work was done.
This work was supported in part by the U.S. Department of Energy Grant No.
DE-FG-02-91ER40685.

\vfill\eject

\noindent {\bf References}

\medskip

\item{1.} L.D.Faddeev and L.A. Takhtajan, \lq\lq Hamiltonian Methods in the
Theory of Solitons", Springer, Berlin,1987.
\item{2.} A. Das, \lq\lq Integrable Models", World Scientific, Singapore,1989.
\item{3.} L.A. Dickey, \lq\lq Soliton Equations and Hamiltonian Systems", World
Scientific, Singapore, 1991.
\item{4.} L.J.F. Broer, Appl. Sci. Res. {\bf 31}, 377 (1975); D.J. Kaup, Prog.
Theor. Phys. {\bf 54}, 396 (1975).
\item{5.} B. A. Kupershmidt, Comm. Math. Phys. {\bf 99}, 51 (1985).
\item{6.} J.C. Brunelli and A. Das, Phys. Lett. {\bf B337}, 303 (1994); {\it
ibid} {\bf B354}, 307 (1995).
\item{7.} J.C. Brunelli and A. Das, hep-th/9408049; hep-th/9505093;
hep-th/9506096.
\item{8.} S. Krivonos, A. Sorin, hep-th/9504084; S. Krivonos, A. Sorin and F.
Toppan, hep-th/9504138; F. Toppan, hep-th/9506133.
\item{9.} J.C. Brunelli, A. Das and W.-J. Huang, Mod. Phys. Lett. {\bf 9A},
2147 (1994).
\item{10.} A. Das and W.-J. Huang, J. Math. Phys. {\bf 33}, 2487 (1992).
\item{11.} I.M. Gelfand and L.A. Dikii, Funct. Anal. Appl. {\bf 10}, 4 (1976).
\item{12.} S. Ghosh and S. Paul, Phys. Lett. {\bf B341}, 293 (1995).
\end